\documentclass[nohyper,11pt,letterpaper]{JHEP3}
\usepackage[dvips]{epsfig}
\usepackage{amsmath}

\newcommand{\be}{\begin{equation}}
\newcommand{\ee}{\end{equation}}
\newcommand{\ben}{\begin{eqnarray}}
\newcommand{\een}{\end{eqnarray}}
\newcommand{\ba}{\begin{eqnarray}}
\newcommand{\ea}{\end{eqnarray}}

\newcommand{\beq}{\begin{equation}}
\newcommand{\eeq}{\end{equation}}

\newcommand{\ie}{{\it i.e.,}\ }

\newcommand{\mt}[1]{\textrm{\tiny #1}}

\newcommand{\seff}{S_\mt{eff}}
\newcommand{\bi}{\begin{itemize}}
\newcommand{\ei}{\end{itemize}}
\newcommand{\ii}{\item}

\newcommand{\lb}{\left (}
\newcommand{\rb}{\right )}
\newcommand{\ltb}{\left [}
\newcommand{\rtb}{\right ]}
\newcommand{\ep}{\epsilon}
\newcommand{\om}{\omega}
\newcommand{\cA}{{\cal A}}
\newcommand{\cB}{{\cal B}}
\newcommand{\ph}{\phi(r,k)}
\newcommand{\php}{\phi'(r,k)}
\newcommand{\cph}{\phi(r,-k)}
\newcommand{\cphp}{\phi'(r,-k)}
\newcommand{\phpp}{\phi''(r,k)}
\newcommand{\cphpp}{\phi''(r,-k)}
\newcommand{\p}{\partial}
\newcommand{\ra}{\rightarrow}
\newcommand{\intk}{\int {d^4 k \over (2 \pi)^4}}
\newcommand{\nt}{{1\over 16 \pi G_5}}
\newcommand{\hhp}{\hspace{.15cm}}

\newcommand{\bc}{\begin{center}}
\newcommand{\ec}{\end{center}}

\newcommand{\hpt}{\hspace{2cm}}
\newcommand{\hpo}{\hspace{1cm}}

\title{Higher Derivative Corrections to Shear Viscosity from
  Graviton's Effective Coupling}

\author{Nabamita Banerjee$^a$ and Suvankar Dutta$^b$\\
$^a$ Harish-Chandra Research Institute, Allahabad, India, \\
$^b$ Dept. of 
Mathematical and 
Statistical Science, University of Alberta, Canada\\
E-mails: $^a$ {\bf nabamita@mri.ernet.in}, \ \ $^b$ {\bf
  sdutta@math.ualberta.ca}}

\abstract{The shear viscosity coefficient of strongly coupled boundary gauge
  theory plasma depends on the horizon value of the effective coupling of
  transverse graviton moving in black hole background. The proof for the
  above statement is based on the canonical form of graviton's action. But in
  presence of generic higher derivative terms in the bulk Lagrangian the
  action is no longer canonical.  We give a procedure to find an effective
  action for graviton (to first order in coefficient of higher derivative
  term) in canonical form in presence of any arbitrary higher derivative terms
  in the bulk. From that effective action we find the effective
  coupling constant for transverse graviton which in general depends on the
  radial coordinate $r$. We also argue that horizon value of this effective
  coupling is related to the shear viscosity coefficient of the boundary fluid
  in higher derivative gravity.  We explicitly check this procedure for two
  specific examples: (1) four derivative action and (2) eight derivative
  action ($Weyl^4$ term). For both cases we show that our results for shear
  viscosity coefficient (upto first order in coefficient of higher derivative
  term) completely agree with the existing results in the literature. }


\preprint{}

\begin{document}{\vskip 1cm}

\section{Introduction} \label{intro}

The AdS/CFT correspondence is a powerful tool to study different properties of
strongly coupled gauge theory in terms of dual (super) gravity theory in AdS
space. In low frequency limit the boundary field theory can be described by
hydrodynamics. In this limit different transport coefficients like shear
viscosity, diffusion constant, thermal and electrical conductivity of strongly
coupled boundary fluid have been computed in the context of AdS/CFT (see
\cite{pss} - \cite{gy}).

In \cite{pss}, the authors evaluated the shear viscosity coefficient of
boundary fluid using Kubo formula.  This formula relates the shear viscosity
to two point function of energy momentum tensor in zero frequency limit. On
the other hand from field operator correspondence of the AdS/CFT conjecture we
know that energy momentum tensor of boundary field theory is sourced by bulk
graviton excitations. Therefore in the context of AdS/CFT, to calculate
thermal two point correlation function of field theory energy momentum tensor
we need to add small perturbations to the bulk metric. In \cite{pss}, the
authors considered graviton excitations polarized parallel to the black brane
(\ie $xy$ components are turned on) and moving transverse to it. When one
sends the gravitons to the brane, there is a probability that it will be
absorbed by the brane. They calculated the absorption coefficient and showed
that it is related to two point functions of energy momentum tensor of
boundary fluid.

To calculate the absorption coefficients, one needs to solve the wave equation
for transverse gravitons.  In presence of any higher derivative terms in the
bulk action the solution may be technically difficult in general
\cite{buchel,kats,myers1}. Recently there is a proposal that the shear
viscosity of strongly coupled boundary gauge theory plasma is related to the
effective coupling of graviton calculated at the black hole horizon
\cite{ramy1,ramy2}. In \cite{liu}, using membrane paradigm, the authors have
confirmed that at the level of linear response the low frequency limit of
strongly coupled boundary field theory at finite temperature is determined by
the horizon geometry of its gravity dual. They have proved that generic
boundary theory transport coefficients can be expressed in terms of geometric
quantities evaluated at the horizon\footnote{See \cite{stretch} also.}. In
particular, they have found that the shear viscosity coefficient is given by
transverse graviton coupling computed at the horizon. The novelty of this
result is that one does not need to solve the equation of motion for the
graviton to calculate the thermal Green function. From graviton's action one
can easily read off the coupling constant and hence determine the shear
viscosity coefficient.

To find the effective coupling of gravitons one has to find the general
action. This can be achieved in the following way. Consider the
Einstein-Hilbert action with negative cosmological constant
\be\label{acn0} 
I = \nt \int d^5x \sqrt{-g}\lb R + 12\rb\ .
\ee 
The equation of motion obtained from this action has a black hole solution. We
denote this background solution by $g^{(0)}_{\mu \nu}$.  Now we consider
fluctuation about this spacetime in $xy$ (for example)
direction\footnote{Notations: $x$ denotes the boundary
coordinates. $x=\{t,{\vec x}\}$.},
\be 
g_{xy}=g_{xy}^{(0)} + \epsilon \ h_{xy}(r,x) = g_{xy}^{(0)} \lb 1
+ \epsilon \ \Phi(r,x)\rb \ .  
\ee 
Then substituting the metric with fluctuation in the action (\ref{acn0}) and
keeping terms up to order $\epsilon^2$ we get the action for graviton. The
form of this action is,
\be \label{0ordform} 
S \sim \nt \intk dr \lb a(r)  \php \cphp + b(r) \ph \cph \rb \ee 
where, 
\be \label{phifu}
\ph = \int {d^4x \over (2 \pi)^4} e^{-i k.x} \Phi(r,x)\ ,
\ee 
$k=\{-\omega,{\vec
k}\}$ and `\ $'$\ ' denotes derivative with respect to $r$.
The effective coupling is related to the coefficient of $\phi^{'2}$
\ie $a$ (we have reviewed this calculation in section \ref{rev}).

This gives the correct viscosity coefficient for the Einstein-Hilbert gravity.
But it is not obvious how to generalize this approach for higher derivative
case.  The proof given in \cite{liu} was based on the canonical form
(\ref{0ordform}) of graviton's action.  In presence of arbitrary higher
derivative terms in the bulk, the general action for the perturbation $h_{xy}$
does not have the above form (\ref{0ordform}). Rather it will have more than
two derivative (with respect to $r$) terms. \cite{liu,cai1} have considered
Gauss-Bonnet term in the bulk action. In general, presence of $R_{ab}R^{ab}$
and $R_{abcd}R^{abcd}$ terms in the bulk result terms like $\phi^{''2}$ and
$\phi' \phi''$ in the action for $h_{xy}$. For Gauss-Bonnet combinations these
terms get canceled and the general action still has the form (\ref{0ordform}).

In this paper we have considered generic higher derivatives terms in the bulk
Lagrangian.  We have given a procedure to construct an effective action
$\seff$ for transverse graviton of the form (\ref{0ordform}) in presence of
any higher derivative terms in the bulk.  The details of the construction is
given in section (\ref{our}).  Our construction ensures that in low frequency
limit, the calculations of retarded Green function (imaginary part) using
either effective action or original action are same.  Therefore following the
similar argument given in \cite{liu}, we can relate the shear viscosity
coefficient of the boundary fluid with the horizon value of the effective
coupling obtained from $\seff$ (section \ref{flow}). In section (\ref{mem}) we
have also discussed how membrane fluid captures the properties of boundary
fluid in low frequency limit in generic higher derivative gravity.  
We have checked our procedure for two cases:
\bi 
\ii 
General four derivative terms, (section (\ref{gbsec}))
\ii 
$Weyl^4$ term which arises in type
II string theory (section (\ref{r4sec})).  
\ei 
In both examples we get exact agreement
between our results and the results that already exist in the literature
\cite{kats,myers1,buchel2}. 
Hence we
conclude that:\\
\\ {\it {The shear
viscosity coefficient of the boundary fluid is given by the horizon value of
the effective coupling of transverse graviton obtained from its effective
action in presence of arbitrary higher derivative terms in the bulk}.}\\


\section{Shear Viscosity from Effective Coupling}\label{rev}

In this section we briefly review how to calculate the shear viscosity
coefficient of the boundary fluid from the effective coupling constant of
transverse graviton in Einstein-Hilbert gravity.  

We  first fix the background spacetime. We start with the
following Einstein-Hilbert action in AdS space. 
\be 
I=\nt \int d^5x \sqrt{-g} \lb R + 12 
\rb\ .
\ee
Here we have taken the radius of the AdS space 1. The background
spacetime is given by the following metric\footnote{We are working 
in a coordinate frame where
  asymptotic boundary is at $r \ra 0$.}
\be
ds^2=-h_t(r)dt^2 + {dr^2 \over h_r(r)} + {1\over r} d{\vec x}^2
\ee 
where,
\be
h_t(r)={1-r^2 \over r} \ .
\ee
and 
\be
h_r(r)=4r^2 (1-r^2) \ . 
\ee 
The black hole has horizon at $r_0=1$ and the temperature of this black hole
is given by,
\be
T={1 \over \pi} \ .
\ee

We 
consider the following metric perturbation,
\be\label{petmet}
g_{xy}=g^{(0)}_{xy}+ h_{xy}(r,x)=g^{(0)}_{xy}(1+\ep \Phi(r,x))
\ee
where $\epsilon$ is an order counting parameter. We  consider terms up to
order $\epsilon^2$ in the action of $\Phi(r,x)$.
The action (in momentum space) is given by,
\ben \label{acn01}
S &=& \nt \int {d\om d^3{\vec k}\over (2 \pi)^4}dr \bigg [{\cA}_{1,1}(r) 
\cphp \php \nonumber \\
&&\hspace{2cm} +
{\cA}_{1,0}(r,k) \cph \php + {\cA}_{0.0}(r,k) \ph \cph \bigg ]
\een
where, $\cA_{i,j}(r,k)$ are functions of r and k and $\ph$ is given by
(\ref{phifu}). 
Up to some total derivative the action (\ref{acn01}) can be written
as\footnote{Though throughout this paper we have written the four vector $k$,
  but in practice we have worked in ${\vec k}\ra 0$ limit. In all 
the expressions we have
  dropped the terms proportional to ${\vec k}$ or its power.},
\be \label{acn02}
S=\nt \int {d\om d^3{\vec k}\over (2 \pi)^4}dr \lb \cA^{(0)}_1(r) \cphp \php
+ \cA^{(0)}_0(r,k) \ph \cph \rb
\ee
where,
\be
\cA_1^{(0)}(r) = {r^2 -1 \over r}
\ee
and 
\be
\cA_0^{(0)}(r,k)= {\omega^2 \over 4r^2(1-r^2)}\ .
\ee
This can be viewed as an action for minimally coupled scalar field $\ph$ with 
effective coupling given by,
\be
K_{\text{eff}}(r)=\nt {\cA^{(0)}_1(r) \over \sqrt{-g^{(0)}} g^{rr}}\ .
\ee
Therefore according to \cite{liu,cai1} the effective coupling 
$K_{\text{eff}}$ calculated at the horizon $r_0$ gives the shear viscosity
coefficient of boundary fluid,
\ben
\eta &=& r_0^{-{3\over 2}} (-2 K_{\text{eff}}(r_0)) \nonumber \\
&=& \nt \ . 
\een


\section {The Effective Action}\label{our}\label{gendis}

Having understood the above procedure to determine the shear viscosity
coefficient from the effective coupling of transverse graviton it is tempting
to generalize this method for any higher derivative gravity. As we discussed
in the introduction, the first problem one faces is that the action for
transverse graviton no more has the canonical form (\ref{acn01}). For
generic 'n' derivative gravity theory the action can have terms with (and up
to) `n' derivatives of $\Phi(r,x)$. Therefore, from that action it is not very
clear how to determine the effective coupling. In this
section we try to address this issue.

We construct an effective action which is of form (\ref{acn02}) with different
coefficients capturing higher derivative effects. We determine these two
coefficients by claiming that the equation of motion for $\ph$ coming from
these two actions (general action and effective action) are same up to first
order in perturbation expansion (in coefficient of higher derivative
term). Once we determine the effective action for transverse graviton in
canonical form then we can extract the effective coupling from the coefficient
of $\php \cphp$ term in the action.  Needless to say, our method is
perturbatively correct.

\subsection {The General Action and Equation of Motion}

Let us start with a generic '$n$' derivative term in the action with
 coefficient $\mu$. We study this system perturbatively and all our
 expressions are valid up to order $\mu$. The action is given by,
\be \label{gacnhd}
S=\nt \int d^5x \lb R + 12 + \mu  \ {\cal R}^{(n)}
\rb
\ee
where, ${\cal R}^{(n)}$ is any $n$ derivative Lagrangian. 
The metric in general is given by (assuming planar symmetry),
\be
ds^2=-(h_t(r)+\mu\ h_t^{(n)}(r))dt^2 + {dr^2 \over h_r(r) + \mu\ h_t^{(n)}(r)} 
+ {1\over r}
(1+ \mu\
h_s^{(n)}(r))d{\vec x}^2 \ 
\ee 
where $h_t^{(n)},h_r^{(n)}$ and $h_s^{(n)}$ are higher derivative corrections
 to the  metric.

Substituting the background metric with fluctuations in the action
(\ref{gacnhd}) (we call it general action or original action) 
for the scalar
field $\ph$ we get,
\be \label{ghdacnphi}
S=\nt \int {d^4 k \over (2 \pi)^4} dr \sum_{p,q=0}^{n} \cA_{p,q}(r,k)
\phi^{(p)}(r,-k) \phi^{(q)}(r,k)
\ee
where, $\phi^{(p)}(r,k)$ denotes the $p^{th}$ derivative of the field $\ph$
with respect to $r$ and $p+q\leq n$. The coefficients $\cA_{p,q}(r,k)$ in
general depends on the coupling constant $\mu$. $\cA_{p,q}$ with $p+q \ge 3$
are proportional to $\mu$ and vanishes in $\mu \ra 0$ limit , since the terms
$\phi^{(p)} \phi^{(q)}$ with $p+q\ge 3$ appears as an effect of higher
derivative terms in the action (\ref{gacnhd}).  Up to some total derivative
terms, the general action
(\ref{ghdacnphi}) can also be written as,
\ben \label{ghdacnphi2}
S&=&\nt \int {d^4 k \over (2 \pi)^4} dr \sum_{p=0}^{n/2}\cA_{p}(r,k)
\phi^{(p)}(r,-k) \phi^{(p)}(r,k), \hspace{.88cm} n \quad \text{even}
 \nonumber \\
&=&\nt \int {d^4 k \over (2 \pi)^4} dr \sum_{p=0}^{(n-1)/2}\cA_{p}(r,k)
\phi^{(p)}(r,-k) \phi^{(p)}(r,k), \hspace{.4cm} n \quad \text{odd} \ .
\een
The equation of motion for the scalar field $\ph$ is given by,
\ben
\sum_{p=0}^{n/2} \lb - {d \over dr} \rb ^{p}{\p {\cal L}(\{\phi^{(m)}\}) 
\over  \p
  \phi^{(p)}(r,k)} =0, \hspace{.88cm} n \quad \text{even} \nonumber \\
\sum_{p=0}^{(n-1)/2} \lb - {d \over dr} \rb ^{p}{\p {\cal L}(\{\phi^{(m)}\}) 
\over  \p
  \phi^{(p)}(r,k)} =0, \hspace{.5cm} n \quad \text{odd}
\een
where ${\cal L}(\{\phi^{(m)}\})$ is given by
\be
{\cal L}(\{\phi^{(m)}\})=\sum_{p}\cA_{p}(r,k)
\phi^{(p)}(r,-k) \phi^{(p)}(r,k)  \ .
\ee

We  analyze the general action
for the scalar field $\ph$ and their equation of motion perturbatively and
write an effective action for the field $\ph$.

The generic form of the equation of motion (varying the general action) 
upto order $\mu$
is given by,
\be\label{geom}
\cA_0(r,k) \ph - \cA_1^{'}(r,k) \php -\cA_1(r,k) \phpp = \mu \ {\hat {\cal
    F}}(\{\phi^{(p)}\}) + {\cal O}(\mu^2) 
\ee
where ${\hat {\cal
    F}}(\{\phi^{(p)}\})$ is some linear function of double and higher 
derivatives of
$\ph$, coming from two or higher
derivative terms in action (\ref{ghdacnphi}).
The zero$^{th}$ order ($\mu \ra 0$) equation of motion is given
by, 
\be \label{eom0}
\cA_0^{(0)}(r,k) \ph - \cA_1^{'(0)}(r,k) \php -\cA_1^{(0)}(r,k) \phpp =0
\ee
where, $\cA_p^{(0)}$ is the value of $\cA_p$ at $\mu \ra 0$. From this
equation we can write $\phpp$ in terms of $\php$ and $\ph$ in $\mu \ra 0$
limit.
\be \label{phi20}
\phpp= {\cA_0^{(0)}(r,k)\over  \cA_1^{(0)}(r,k)} \ph - {\cA_1^{'(0)}(r,k)
  \over \cA_1^{(0)}(r,k)} \php \ .
\ee
Then the full equation of motion can be written in the following way,
\ben \label{eomg}
\cA_0^{(0)}(r,k) \ph - \cA_1^{'(0)}(r,k) \php -\cA_1^{(0)}(r,k) \phpp &=& \mu\ 
   {\tilde {\cal F}}(\ph,\php,\phpp,...) \nonumber \\
&&\hspace{2cm}  + {\cal O}(\mu^2) \ . 
\een
Since the right hand side of equation (\ref{eomg}) is proportional to $\mu$, we
can replace the $\phpp$ and other higher (greater than 2) derivatives of $\ph$
by its leading order value (\ref{phi20}). Therefore up to order $\mu$ the
equation of motion for $\phi$ is given by,
\ben \label{eomg2}
\cA_0^{(0)}(r,k) \ph - \cA_1^{'(0)}(r,k) \php -\cA_1^{(0)}(r,k) \phpp &=& \mu\ 
   {\cal F}(\ph,\php) \nonumber \\
&&\hspace{2cm} + {\cal O}(\mu^2) \nonumber \\
&=& \mu ({\cal F}_1 \php + {\cal F}_0 \ph) \nonumber \\
&& \hspace{2cm} + {\cal O}(\mu^2) \ 
\een
where ${\cal F}_0$ and ${\cal F}_1$ are some function of $r$.
This is the perturbative equation of motion for the scalar field $\ph$
obtained from the general action (\ref{ghdacnphi}).

\subsection{Strategy to Find The Effective Action}\label{gendis1}

In this subsection we describe the strategy to write an effective action for
the field $\ph$ which has form (\ref{acn02}) with different functions. 
The prescription is
following.
\bi 
\ii 
{\bf (a)} We demand the equation of motion for $\ph$ obtained from the
original action and the effective action are same upto order $\mu$. This will
fix the coefficients of $\phi^{'2}$ and $\phi^2$ terms in effective action.

Let us start with the following form of the effective action.
\ben \label{acneff}
\seff&=& {1 \over 16 \pi G_5} 
\int {d\om d^3{\vec k}\over (2 \pi)^4}dr \bigg[ (\cA^{(0)}_1(r,k) +
\mu \cB_1(r,k)) \cphp \php \nonumber \\
&&\hspace{2cm} + (\cA^{(0)}_0(r,k)+\mu \cB_0(r,k)) \ph \cph \bigg ]\ .
\een
The functions $\cB_0$ and $\cB_1$ are yet to be determined. We  determine
these functions by claiming that the equation of motion for the scalar field
$\ph$ obtained from this effective action is same as (\ref{eomg2}) 
up to order $\mu$. 
The equation of motion for $\ph$ from the effective action is given by,
\ben
\cA_0^{(0)}(r,k) \ph &-& \cA_1^{'(0)}(r,k) \php -\cA_1^{(0)}(r,k) \phpp 
\nonumber \\
&=& \mu \lb \cB_1^{'}(r,k) - {\cA_1^{'(0)}(r,k)
  \over \cA_1^{(0)}(r,k)} \cB_1(r,k) \rb \php  \nonumber\\ 
&& + \mu \lb \cB_1(r,k) {\cA_0^{(0)}(r,k)\over
  \cA_1^{(0)}(r,k)} - \cB_0(r,k) \rb \ph + {\cal O}(\mu^2) \ .
\een
Therefore comparing with (\ref{eomg2}) we get,
\be
\cB_1'(r,k)  - {\cA_1^{'(0)}(r,k)
  \over \cA_1^{(0)}(r,k)} \cB_1(r,k) - {\cal F}_1(r,k)  =0
\ee 
and 
\be\label{cb00}
\cB_0(r,k)= \cB_1(r,k) {\cA_0^{(0)}(r,k)\over
  \cA_1^{(0)}(r,k)} - {\cal F}_0(r,k) \ .
\ee
The solutions are given by,
\ben \label{cb1}
\cB_1(r,k) &=& \cA_1^{(0)}(r,k)\int dr {{\cal F}_1(r,k) 
\over \cA_1^{(0)}(r,k)} +
\kappa \cA_1^{(0)}(r,k) 
\nonumber \\
&=&{\tilde {\cal B}}_1(r,k) + \kappa  \cA_1^{(0)}(r,k) 
\een
and
\be\label{cb0}
\cB_0= {\tilde \cB}_0(r,k) + \kappa \cA_0^{(0)}\ 
\ee
for some constant $\kappa$. We need to fix this constant.

\ii 
{\bf (b)} Condition {\bf (a)} 
can not fix the overall normalization 
factor of the
effective action.  
In particular we can multiply it by $(1 +
\mu \Gamma)$ (for some constant $\Gamma$) and still get the same equation of
motion\footnote{We are thankful to
  Ashoke Sen for raising this 
  point.}. 
Considering this normalization, the effective action is given by,
\ben
\seff&=& {\ 1 + \mu \ \Gamma \ \over 16 \pi G_5} 
\int {d\om d^3{\vec k}\over (2 \pi)^4}dr \bigg[ (\cA^{(0)}_1(r,k) +
\mu \cB_1(r,k)) \cphp \php \nonumber \\
&&\hspace{2cm} + (\cA^{(0)}_0(r,k)+\mu \cB_0(r,k)) \ph \cph \bigg ]\ .
\een
Substituting the values of $\cB$'s (\ref{cb1}) and (\ref{cb0}) we get,
\be
\seff=(1+ \mu (\Gamma+\kappa)) S^{(0)} + \mu\int dr \lb {\tilde {\cal
    B}}_1(r,k) \cphp \php + {\tilde \cB}_0(r,k) \cph \ph \rb
\ee
where $S^{(0)}$ is the effective action at $\mu \ra 0$ limit. This implies
that the integration constant $\kappa$ can be absorbed in 
the overall normalization
constant $\Gamma$. Henceforth we will denote this combination as $\Gamma$.
 
Our prescription is to take $\Gamma$ to be ${\bf zero}$ from the following
observation. 
\bi
\ii
The shear viscosity coefficient of boundary fluid is related to the imaginary
part of retarded Green function in low frequency limit. The retarded 
Green function $G^R_{xy,xy}(k)$ 
is defined in the following way. The on-shell action for
graviton can
be written as a surface term,
\ben \label{rgf} 
S \sim  \intk
\,\phi_0(k)\,{\cal G}_{xy,xy}(k,r)\,\phi_0(-k)
\een
where $\phi_0(k)$ is the boundary value of $\ph$ 
and $G^{R}_{xy,xy}$ is given by,
\ben
G^R_{xy,xy}(k)=\lim_{r\ra 0}2 {\cal G}_{xy,xy}(k,r)
\een
and shear viscosity coefficient is given by\footnote{To calculate this number
  one has to know the exact solution, \ie the form of $\xi$ and the value of
  $\beta$ in (\ref{phiform}).},
\be
\eta = \lim_{\omega \ra 0} \bigg[ {1\over \omega} \text{Im} G^R_{xy,xy}(k)
  \bigg ] \qquad (computed \ \ on-shell)\ .
\ee
\ii
Now it turns out that the imaginary part of this retarded Green function
obtained from the original action and effective action are same upto the
normalization constant $\Gamma$ in presence of generic higher derivative terms
in the bulk action. Therefore it is quite natural to take $\Gamma$ to be
$zero$ as it ensures that starting from the effective action also one can get
same shear viscosity using Kubo machinery. To show that the above statement is
true we do not need to know the full solution for $\phi$, in other words to
find the difference between the two Green functions one does not 
need to calculate the
Green functions explicitly. Assuming the following general form of
solution for $\phi$
\be \label{phiform}
\phi \sim (1-r^2)^{-i \omega \beta} \lb 1 + i \omega \beta \mu \xi(r) \rb
\ee
it can be shown generically.  In appendix \ref{app1} we have given the
proof.
\ii
Because of the canonical form of the effective action, it follows from the
argument in \cite{liu} and the statement above, that the shear viscosity
coefficient of boundary fluid is given by the horizon value of the effective
coupling obtained from the effective action in presence of any higher
derivative terms in the bulk action. We discuss elaborately on this point in
section (\ref{flow}). 
\ei
\ii
{\bf (c)} After getting the effective action for 
$\ph$, the effective coupling is given by,
\be\label{gkeff}
K_{\text{eff}}(r)=  {1\over 16 \pi G_5}\ 
{\cA_1^{(0)}(r,k) + \mu \cB_1(r,k) \over 
\sqrt{-g}
  g^{rr}}
\ee 
where $g^{rr}$ is the '$rr$' component of the inverse perturbed metric and 
$\sqrt{-g}$ is the determinant of the perturbed metric. Hence the shear
viscosity coefficient is given by,
\be
\eta = r_0^{-{3\over 2}} (-2 K_{\text{eff}}(r=r_0))
\ee
where $r_0$ is the corrected horizon radius.
\ei
To summaries, we have obtained a well defined procedure to find the correction
(up to order $\mu$) to the coefficient of shear viscosity of the boundary
fluid in presence of
general higher derivative terms in the action.



\section{Flow from Boundary to Horizon}\label{flow}

Following \cite{liu}, let us 
define the following linear response
function
\be\label{defchi}
{\bar \chi}(r,k) = {\Pi(r,k) \over i \omega \ph}
\ee
where $\Pi(r,k)$ is conjugate momentum of the scalar field
$\phi$ (with respect to a foliation in the $r$ direction),
\ben\label{pidef}
\Pi(r,k) &=& \lb \cA_1^{(0)}(r,k) + \mu \cB_1(r,k) \rb \cphp \nonumber \\
&=& {\tilde K_{\text{eff}}}(r) \sqrt{-g^{(0)}} g^{(0)^{rr}} \p_r \phi
\een
where ${\tilde K_{\text{eff}}}(r) = 16 \pi G_5 K_{\text{eff}}(r)$.
Now we will show, using the equation of motion,
that the function $\Pi(r,k)$ and the combination $\omega \phi(r,k)$ 
is independent of the radial
coordinate $r$ in $k\ra 0$ limit. The equation of motion is given by,
\ben \label{floweq}
{d \over dr} \bigg [\lb \cA_1^{(0)}(r,k) + \mu \cB_1(r,k) \rb \php\bigg ] 
= \lb
\cA_0^{(0)}(r,k) + \mu \cB_0(r,k) \rb \ph \nonumber \\
{d \over dr}\bigg [ \Pi(r,k)\bigg ] = \lb
\cA_0^{(0)}(r,k) + \mu \cB_0(r,k) \rb \ph \ .
\een
Since $\cA_0^{(0)} \sim \omega^2$, therefore it follows from (\ref{floweq})
and (\ref{pidef}) that, in
$\mu \ra 0$ limit  $\Pi(r,k)$ and $\omega \ph$ 
are independent of $r$. But this is true even in
$\mu \neq 0$ case. To understand this we note that, function $\cA_0$ in
(\ref{ghdacnphi2}) is proportional to 
$\omega^2$ in general\footnote{In general when we write action
  (\ref{ghdacnphi2}) action (\ref{ghdacnphi}) we get some terms like
 $\omega^2 \phi^2 + Z(r) \phi^2$. The function $Z(r)$ is zero when background
  equation of motion is satisfied. We have explicitly checked this for two,
  four and eight derivative case.}.  Therefore it follows from
(\ref{phi20}), (\ref{eomg2}) and (\ref{cb00}) that $\cB_0$ is also
proportional to $\omega^2$. Hence, in presence of higher derivative terms
also it follows from (\ref{pidef}) and (\ref{floweq}) that 
the function $\Pi(r,k)$ and $\omega \phi(r,k)$ are 
independent of radial direction $r$ in low
frequency limit.

Therefore this response 
function calculated at the asymptotic
boundary and at the horizon gives the same result and is equal to the shear
viscosity coefficient.
One can calculate the function $\tilde {\chi}$ 
and it turns
out that,
\ben
{\bar \chi}(r=0,k\ra 0) &=& {\text{Im}G^{R^{\text{eff}}}_{xy,xy} \over i 
\omega}\ , \nonumber \\
{\bar \chi}(r=r_0,k\ra 0) &=& -{r_0^{-3/2}\over 8 \pi G_5}\ 
{\cA_1^{(0)}(r,k) + \mu \cB_1(r,k) \over 
\sqrt{-g}
  g^{rr}}\bigg |_{r_0}=r_0^{-{3\over 2}} \lb -2 K_{\text{eff}}(r_0)\rb \ .
\een
Thus, shear viscosity coefficient of boundary fluid is related to
horizon value of graviton's effective coupling obtained from the effective
action.


\section{Membrane Fluid in Higher Derivative Gravity}\label{mem}

The {\bf UV/IR} connection tells us that the boundary field theory physics in
low frequency limit should be governed by the near horizon geometry of its
gravity dual. In \cite{liu}, the authors have established a connection between
horizon membrane fluid and boundary fluid in linear response
approximation. They considered a mass less scalar field (with action given in
(\ref{acn02})) outside the horizon and studied the response of the membrane
fluid to this bulk scalar field. They defined a {\it membrane charge}
$\Pi_{\text{mb}}$ which is equal to the conjugate momentum of the scalar field
$\phi$ (with respect to a foliation in the $r$ direction) at the
horizon. Imposing regularity condition on the scalar field at the horizon they
interpreted the membrane charge $\Pi_{\text{mb}}$ as a response of the horizon
fluid to the scalar field. Considering the scalar field $\phi$ to be bulk
graviton excitation ($h_{x}^y$), $\Pi_{\text{mb}}$ gives the shear viscosity
of the membrane (horizon) fluid which is also equal to horizon value of the
effective coupling of graviton. In this way, they proved that the shear
viscosity of boundary fluid is related to that of membrane fluid.

In higher derivative gravity, since the canonical form of the action
(\ref{acn02}) breaks down, it is not very obvious how to define the membrane
charge $\Pi_{\text{mb}}$. Instead of the original action if we consider the
effective action (\ref{acneff}) for graviton then it is possible to write the
membrane action perturbatively and define the membrane charge
($\Pi_{\text{mb}}$) in higher derivative gravity. As if the membrane fluid is
sensitive to the effective action $\seff$ in higher derivative gravity.

Following \cite{liu} we can write the membrane action and charge in the
following way (in momentum space)
\be
S_{\text{mb}}= \int_{\Sigma} {d^4k \over (2 \pi)^4}
 \sqrt{-\sigma} \lb{\Pi(r_0,k) \over
  \sqrt{-\sigma}} \phi(r_0,-k) \rb
\ee 
where $\sigma_{\mu\nu}$ is the induced metric on the membrane and 
$\Pi(r,k)$ is given by (\ref{pidef})
and the membrane charge is given by,
\be
\Pi_{\text{mb}} = {\Pi(r_0,k) \over \sqrt{-\sigma}}=- {\tilde
  K}_{\text{eff}}(r_0) \sqrt{g^{(0)^{rr}}} \p_r
\ph \big |_{r_0} \ .
\ee
Imposing the in-falling wave boundary condition on $\phi$, it can be shown
that the membrane charge $\Pi_{\text{mb}}$ is the response of the horizon
fluid to the bulk graviton excitation and the membrane fluid transport
coefficient is given by,
\be
\eta_{\text{mb}}= {\tilde K}_{\text{eff}}(r_0) \ .
\ee

Hence, we see that even in higher derivative gravity the shear viscosity
coefficient of boundary fluid is captured by the membrane fluid.



\section{Four Derivative Lagrangian} \label{gbsec}

In this section we  apply our effective action approach to calculate the
correction to the shear viscosity in presence of general four derivative terms
in the action. The four derivative bulk action we  consider is of the
following form
\be \label{4acn}
S=\nt \int d^5x  \ltb R + 12 + \mu \lb c_1 R^2 + c_2 R_{ab}R^{ab} + c_3
R_{abcd} R^{abcd} \rb \rtb
\ee
with constant $c_1$, $c_2$ and $c_3$.
The background metric is given by,
\ben\label{gbmet1}
ds^2= -{f(r) \over r} dt^2 + {dr^2 \over 4 r^2 f(r)} + {1 \over r} d{\vec x}^2 
\een
where,
\be
f(r)= 1-r^2 + {\mu \over 3}(4 (5c_1 + c_2) + 2c_3)  + 2 \mu c_3 r^4 \ .
\ee
The position of the horizon is given by,
\be
f(r_0) =0
\ee
which implies that,
\be
r_0= 1+ {2\over 3} (5c_1 + c_2 + 2 c_3) \mu + {\cal O}(\mu^2) \ .
\ee
The temperature of this black hole is given by,
\be
T=
\frac {1} {\pi } + \frac {(5 {c_1} +  {c_2} - 
      7  {c_3}) \mu} {3 \pi
      } + O\left (\mu^2 \right) \ .
\ee
In this coordinate frame the boundary metric is given by,
\be
ds_4^2= \lb -f(0) dt^2 + d{\vec x}^2 \rb
\ee
which is not Minkowskian. Therefore we  rescale our time coordinate to make
the boundary metric Minkowskian. We replace,
\be
t \ra {t \over \sqrt{f(0)}} 
\ee
in the metric (\ref{gbmet1}). The rescaled metric is,
\ben \label{gbmet2}
ds^2= -{f(r) \over f(0)  r} dt^2 + {dr^2 \over 4 r^2 f(r)} + {1 \over r}
d{\vec x}^2 \ .
\een
This is our background metric and we consider fluctuation around this. 

\subsection{The General Action}

In this theory, the general action for the scalar field $\ph$ is given by,
\ben
S&=&\nt \int {d^4 k \over (2 \pi)^4} dr \bigg [A^{GB}_1(r,k) 
\ph \cph + A_2^{GB}(r,k) \php
\cphp \nonumber \\
&& \hspace {2cm}
+ A_3^{GB}(r,k) \phpp \cphpp + A_4^{GB}(r,k) \ph \cphp \nonumber \\
&& \hspace{2cm}+ A_5^{GB}(r,k) \ph \cphpp 
+ A_6^{GB}(r,k)
\php \cphpp \bigg ]
\een
where the expressions for $A^{GB}_i$s are given in appendix \ref{app2}. 
Up to some total
derivative terms this action can be written as,
\ben
S= \nt \int {d^4 k \over (2 \pi)^4} dr \bigg [ \cA_0^{GB} \ph \cph + \cA_1^{GB}
\php \cphp + \cA_2^{GB} \phpp \cphpp \bigg ] \nonumber \\
\een
where,
\ben
\cA_0^{GB} &=& A^{GB}_1(r,k) -{A^{'GB}_4(r,k) \over 2} + {A_5^{''GB}(r,k) 
\over 2}
\nonumber \\
\cA_1^{GB} &=& A_2^{GB}(r,k) - A_5^{GB}(r,k) -{A^{'GB}_6(r,k) \over 2} 
\nonumber \\
\cA_2^{GB} &=& A_3^{GB}(r,k) \ .
\een

\subsection{The Effective Action and Shear Viscosity}

Following the general discussion of section 
(\ref{gendis}) we write
the effective action for the scalar field,
\ben \label{effgb}
\seff^{GB} &=& {(1+ \Gamma \mu) \over 16 \pi G_5} \intk \bigg [
 (\cA^{(0)}_1(r,k) +
\mu \cB_1^{GB}(r,k)) \cphp \php \\  
&& \hspace{2cm} 
+ (\cA^{(0)}_0(r,k)+\mu \cB_0^{GB}(r,k)) \ph \cph \bigg ]\ .
\een
To evaluate the functions $\cB_1^{GB}$ and $\cB_0^{GB}$ and to fix the
normalization constant $\Gamma$, we follow the strategy given in section
(\ref{gendis1}). Comparing the equation of 
motion for $\ph$ from two actions we get 
the function $\cB_1^{GB}$ and $\cB_0^{GB}$ of the following form,
\ben
\cB_0^{GB}&=& {\omega^2 \over 12r^2 (1-r^2)^2} \lb 10 (11r^2 -13) c_1 + (22
r^2 -26) c_2 + (11 -25 r^2 + 6r^4) c_3 \rb \nonumber \\
\cB_1^{GB}&=&  {1\over 3r} \lb (110 -130 r^2)c_1 +(22 -26 r^2 )c_2 -(13 -23
r^2  +
18 r^4)c_3 \rb
\ .
\een
The normalization constant $\Gamma=0$ (appendix \ref{app1}).

Now we can calculate the effective coupling using the formula (\ref{gkeff}). 
It turns out to be,
\be
K_{\text{eff}}(r)= \nt \lb -{1\over 2} + \lb 4(5 c_1 + c_2) - 2(1-r^2)c_3 \rb 
\mu \rb \ .
\ee
Therefore the shear viscosity is given by,
\ben
\eta &=& {1\over r_0^{3/2}} \lb -2 K_{\text{eff}}(r_0) \rb \nonumber \\
&=& \nt {1\over r_0^{3/2}} \lb 1 - 8(5 c_1 + c_2)\mu \rb  \nonumber \\
&=& \nt \lb 1- 9 \mu \ (5c_1 + c_2) - 2 \mu \ c_3 \rb\ .
\een
This result is in agreement with \cite{kats,myers1,dutta}.


\section{String Theory Correction to Shear Viscosity} \label{r4sec}

In this section we  apply the effective action approach for eight
derivative terms in the Lagrangian. We  consider the well known $Weyl^4$
term. This term appears in type II string theory. The five dimensional bulk
action is given by,
\be \label{w4acn}
S=\nt \int d^5x \sqrt{-g} \lb R + 12 + \mu W^{(4)} \rb
\ee
where,
\be
 W^{(4)}=C^{hmnk}C_{pmnq}C_h^{\hhp rsp}C^q_{\hhp rsk}+{1\over
2}C^{hkmn}C_{pqmn}C_h^{\hhp rsp}C^q_{\hhp rsk}
\ee
and the weyl tensors $C_{abcd}$ are given by,
\be
C_{abcd}=R_{abcd} + {1 \over 3}
(g_{ad}R_{cb}+g_{bc}R_{ad}-g_{ac}R_{db}-g_{bd}R_{ca})+{1 \over
  12}(g_{ac}g_{bd}-g_{ad}g_{cb})R\ .
\ee
The background metric is given by \cite{gks,dg},
\ben\label{w4met}
ds^2 &=& -\frac {(1-r^2)}{r}  \left (1 + 45 \mu  r^6 - 75 \mu  r^4 - 
     75 \mu 
         r^2 \right) dt^2 \nonumber \\
&+& {1\over 4 (1-r^2) r^2 } \lb 1-285 \mu  r^6 + 75 \mu  r^4 + 75 \mu  r^2  
   \rb  dr^2 + {1\over r} d{\vec x}^2\ .
\een
The temperature of this black hole is given by,
\be
T= {1 \over \pi} \lb 1 + 15 \mu \rb \ .
\ee
The horizon is located at $r_0=1$.

\subsection{The General Action}

Putting the perturbed metric in (\ref{w4acn}) we get the
general  action for the scalar field $\ph$,
\ben
S&=&\nt \int {d^4 k \over (2 \pi)^4} dr \bigg [
A_1^W(r,k) \ph \cph + A^W_2(r,k) \php
\cphp \nonumber \\
&& \hspace{2cm}+ A_3^W(r,k) \phpp \cphpp + A^W_4(r,k) \ph \cphp 
 \nonumber
\\ 
&&\hspace{2cm}+ A^W_5(r,k) \ph \cphpp + A^W_6(r,k)
\php \cphpp \bigg ] \ .
\een
The coefficients $A_i^W$s are given in appendix (\ref{app4}). Like four
derivative case,
up to some total
derivative terms this action can be written as,
\ben
S= \nt \int {d^4 k \over (2 \pi)^4} dr \bigg [ \cA_0^{W} \ph \cph + \cA_1^{W}
\php \cphp + \cA_2^{W} \phpp \cphpp \bigg] \nonumber \\
\een
where,
\ben
\cA_0^{W} &=& A^W_1(r,k) -{A^{'W}_4(r,k) \over 2} + {A_5^{''W}(r,k) \over 2}
\nonumber \\
\cA_1^{W} &=& A_2^W(r,k) - A_5^W(r,k) -{A^{'W}_6(r,k) \over 2} \nonumber \\
\cA_2^{W} &=& A_3^W(r,k) \ .
\een 

\subsection{The Effective Action and Shear Viscosity}

We  write
the effective action for the scalar field in the following way,
\ben
\seff^{W} &=& {(1+ \Gamma \mu) \over 16 \pi G_5} \intk  \bigg[
 (\cA^{(0)}_1(r,k) +
\mu \cB_1^{W}(r,k)) \cphp \php \\  
&& \hspace{2cm}+ (\cA^{(0)}_0(r,k)+\mu \cB_0^{W}(r,k)) \ph \cph \bigg]\ .
\een
The functions $\cB_0^W$ and $\cB_1^W$ are given by,
\be
\cB_0^W(r,k)=-\frac {\omega ^2 \left (663 r^6 - 573 r^4 + 75 r^2 
     \right)} {4 r^2
       \left (r^2 - 1 \right)}
\ee
\be
\cB_1^W(r,k)=\frac {\left (r^2 - 1 \right) \left (129 r^6 + 141 r^4 - 75 r^2 
\right)} {r} \ .
\ee
The normalization constant $\Gamma=0$ (Appendix \ref{app1}).

The effective coupling constant is given by (\ref{gkeff}),
\ben
K_{\text{eff}}(r) &=& \nt {\cA_1^{(0)}(r,k) + \mu \ \cB_1^W(r,k) 
\over \sqrt{-g}
  g^{rr}}\nonumber \\
&=&\nt \lb -{1\over 2} \lb 1 + 36 \mu \ r^4 (6 -r^2) \rb \rb \ .
\een
Therefore the shear viscosity is given by,
\ben\label{w4eta}
\eta &=&  r_0^{-\frac{3}{2}} (-2 K_{\text{eff}}(r_0)) \nonumber \\
&=& \nt \lb 1 + 180 \ \mu \rb, \ \ \ \ (r_0=1) \ 
\een
and shear viscosity to
entropy density ratio 
\be 
{\eta \over s}={1\over 4 \pi }\lb 1 + 120 \ \mu \rb
\ee 
where entropy density $s$ is given by \cite{gks,dg}, 
\be 
s={1\over 4G_5}
\lb 1 + 60 \ \mu \rb \ .  
\ee 
These results agree with
\cite{buchel2}\footnote{In fact, in \cite{buchel} the result for $\eta/s$ was
not correct. Later the author(s) corrected their results in \cite{buchel2}.}.

\section{Discussion}\label{dis}

We have found a procedure to construct an effective action for transverse
graviton in canonical form in presence of any higher derivative terms in 
bulk and showed that the horizon value of the effective coupling
obtained from the effective action gives the shear viscosity coefficient of
boundary fluid. Our results are valid upto first order in $\mu$. We discussed
two non trivial examples to check the method. We have considered four
derivative and eight derivative ($Weyl^4$) Lagrangian and calculated the
correction to the shear viscosity using our method. We found complete
agreement between our result and the results obtained using other methods.

Since the equation of motion for scalar field $\ph$ obtained from effective
and original actions are same, these two actions should be related by some
field re-definition. If one finds such field re-definition then the
normalization of the effective action will be fixed automatically\footnote{We
  are thankful to Ashoke Sen for discussion on thins point.}.

In \cite{ramy1} the authors have proposed a formula for shear viscosity for
generalized higher derivative gravity in terms of some geometric quantity
evaluated at the event horizon (like Wald's formula for entropy). Though their
proposal gives correct results for Einstein-Hilbert and Gauss-Bonnet action
but unfortunately we are unable to get the correct result for $Weyl^4$ term. 
We find the shear viscosity coefficient for $Weyl^4$ term 
(using their proposal)
\be 
\eta = \nt \lb 1 + 20 \mu
\rb 
\ee 
which implies, 
\be 
{\eta \over s} ={1\over 4 \pi} \lb 1 - 40 \mu \rb\
.  
\ee 
These issues are under investigation \cite{wip}.

In this paper we have concentrated on a particular transport coefficient,
namely the shear viscosity coefficient. But the other transport coefficients
like electrical and thermal conductivity of boundary fluid can also be
captured in terms of membrane fluid. It would also be interesting to study
these other transport coefficients in the context of higher derivative
gravity.

\bc
--------------------------------
\ec

\noindent
{\bf Acknowledgment}

\vspace{.4cm} We would like to thank Rajesh Gopakumar, Dileep Jatkar and
Ashoke Sen for many useful discussions and their comments on our first
draft. We are also thankful to R.G. Cai, D. Gorbons, Y. Kats and V. Suneeta
for communications. SD would like to acknowledge NSERC of Canada. We are
thankful to people of India for their kind support to fundamental research.
\bc * \ \ \ * \ \ \ * \ \ \ * \ \ \ * \ \ \ * \ec

\newpage

\noindent
{\Large {\bf Appendix}}
\appendix

\section{Fixing the Normalization Constant}\label{app1}

In this appendix we fix the normalization constant $\Gamma$. We
consider a general class of action for $\phi$ which appears when the higher
derivative terms are made of different contraction of Ricci tensor, Riemann
tensor, Weyl tensor, Ricci scalar etc. or their different powers. Since, all
these tensors involve two derivatives of metric they can only have terms like 
$\p_a \p_b \Phi(r,x)$ and its lower derivatives. Therefor the most generic
quadratic (in $\Phi(r,x)$, 
in linear response theory) action for this kind of higher derivative gravity 
has the following form (in
momentum space)\footnote{In all the expressions we have omitted $k$ dependence
  of $\phi$.}  
\ben
S&=&\nt \intk dr \bigg[
 \text {a1}(r) \phi (r)^2+ \text{a2} (r)  \phi ' (r)^2 + 
\text{a4}(r) \phi(r)
   \phi'(r) \nonumber \\
&&\hspace{2cm}+ \mu \ \text{a6}(r) \phi''(r) \phi'(r) 
+ \mu \ \text{a3}(r) \phi''(r)^2 + \text{a5}(r) \phi(r) \phi''(r)\bigg]
\een
where,
\ben
a1(r)&=&
\frac {-8 r^2 + \omega ^2 r + 8} {4 r^3 - 4 r^5} + \mu \ 
\text {f2} (r) \nonumber
\\ 
a2(r)&=&
-3 r + \frac {3} {r} + \mu \ \text {h2} (r)  \nonumber \\
a4(r)&=&
- \frac {6} {r^2} - 2 + \mu \ \text {g2} (r) \nonumber \\
a5(r)&=&
-4 r + \frac {4} {r}  + \mu \ \text {j2} (r) 
\een
and $a3(r), a6(r), j2(r), g2(r), h2(r)$ and $f2(r)$ depends on 
higher derivative terms in the action.
Now let us write the effective Lagrangian as follows, 
\ben 
\seff&=& {1 + \mu
\Gamma \over 16 \pi G_5} \intk dr \bigg[
\frac {4 r \left (r^2 - 1 \right)^2
\phi'(r)^2 - \omega^2 \phi(r)^2} {4 r^2 \left (r^2 - 1 \right)}
\nonumber \\
&& \hpt + \mu \bigg ( 
\text {b2}
(r) \phi(r)^2 + \ \text {b1} (r) \phi'(r)^2 \bigg )\bigg] \ .
\een 
From condition {\bf (a)} of
section (\ref{gendis}) the solutions for $b1$ and $b2$ are given by, 
\ben
b1(r)&=&{1 \over 2 r \left (r^2 - 1 \right)^2} (\left (-4 r^3 - 12 r + \omega
^2 \right) \text {a3} (r) 
\nonumber \\
&& + \left (r^2 - 1 \right) 
(2 \kappa
r^4 - \text {a6}'(r) r^3 - 4 \kappa r^2 + 2 \text{a3}'(r) r^2 \nonumber \\
&& + 2 \left (r^2 -
1 \right) \text {h2}(r) r - 2 \left (r^2 - 1 \right) \text{j2}(r) r + \text
{a6}' (r) r + 2 \kappa + 2 \text {a3}' (r))) 
\een
\ben
b2(r)&=&
-{1\over 16 r^2 \left (r^2 - 
      1 \right)^4} (\left (\omega ^4 + 144 r^3 \omega^2 \right) 
\text {a3} (r) \nonumber \\
&& \hpo + 4
        \left (r^2 - 1 \right) (-4 r^2 \text {f2} (r)
           \left (r^2 - 1 \right)^3 + (2 r^2 \text {g2}' (r)
               \left (r^2 - 1 \right)^2 \nonumber \\
&& \hpo + \left (\omega^2 \kappa - 
              2 r^2 \left (r^2 - 1 \right)
                   \text {j2}'' (r) \right) \left (r^2 - 1 \right) \nonumber
		   \\ 
&& \hpo +
           r \omega^2
               \text {a3}'' (r) ) \left (r^2 - 
         1 \right) + \left (1 - 11 r^2 \right) 
\omega^2 \text {a3}' (r) ))\ .
\een
The boundary term coming from the original action (after adding {\it
  Gibbons-Hawking} boundary terms) are given by, 
\ben
S^{{\cal B}}&=& \nt \intk\bigg[
-\frac {\phi (r)^2} {r^2} + \phi (r)^2 + 
 r \phi ' (r) \phi (r) - \frac {\phi ' (r)
        \phi (r)} {r} \nonumber \\
&& \hpt + 
 \mu \bigg (\frac {1} {2} \text {g2} (r) \phi (r)^2 - \frac {1} {2}
         \text {j2}' (r) \phi (r)^2 \nonumber \\
&& \hpt + \text {h2} (r) \phi ' (r) \phi (r) - \text {j2} (r)
         \phi ' (r) \phi (r) - \frac {1} {2} 
\text {a6}' (r) \phi ' (r) \phi
        (r) \nonumber \\
&& \hpt +\frac {
\text {a3}' (r) \left (\phi (r) \omega^2 + 
          4 \left (r^4 - 1 \right)
               \phi ' (r) \right) \phi (r)} {4 r \left (r^2 - 
          1 \right)^2} \nonumber \\
&& \hpt - \frac {\text {a3} (r)
           \left (6 r \phi (r) \omega^2 + \left (r^2 - 
              1 \right) \left (8 r^3 + 24 r - 
\omega^2 \right) \phi ' (r) \right) \phi (r)} {4 r
           \left (r^2 - 1 \right)^3} \nonumber \\
&& \hpt - \frac 
{\text {a3} (r) \phi ' (r) \left (\phi (r)
              \omega^2 + 
         4 \left (r^4 - 1 \right) \phi ' (r) \right)} {4 r
           \left (r^2 - 1 \right)^2} \nonumber \\
&& \hpt + \text {a3} (r) \phi ' (r) \left (-\frac {\phi (r)
              \omega^2} {2 r \left (r^2 - 
             1 \right)^2} - \frac {\left (r^4 - 1 \right) \phi
               ' (r)} {r \left (r^2 - 1 \right)^2} \right) \bigg )\bigg]\ .
\een
And the boundary terms coming from the effective action are given by,
\ben
\seff^{{\cal B}}&=& {1 \over 16 \pi G_5} \intk \bigg[
\left (r - \frac {1} {r} 
\right) \phi (r) \phi ' (r) \nonumber \\
&& \hpt + {\mu \over 2 r \left (r^2 - 1 \right)^2} \bigg (\ \phi (r) (2
            \Gamma  \left (r^2 - 
           1 \right)^3 
+ (- \text {a6}' (r) r^3 
\nonumber \\
&& \hpt + 2 \text {a3}' (r) r^2 + 
           2 \left (r^2 - 1 \right) \text {h2} (r) r - 2
                \left (r^2 - 1 \right) \text {j2} (r) r 
\nonumber \\
&& \hpt + \text {a6}' (r) r  + 2
                \text {a3}' (r) ) \left (r^2 - 
          1 \right) + \left (-4 r^3 - 12 r + 
\omega^2 \right) \text {a3} (r) 
) \phi ' (r)\bigg ) \bigg]\ .
\een
Let the form of the solution of $\phi$ is given by,
\be
\left (1 - r^2 \right)^{i \beta  \omega } \lb 1 + i \beta \omega \mu F(r) \rb
\ee
with
\be
F(0)=0.
\ee
The imaginary part of the retarded Green function for original action is given
  by, 
\ben
{1 \over  \omega} \text{Im}[G^{R^{(\text{original})}}_{xy,xy}]&=& 
\lim_{r\ra 0}\bigg[- 2 \beta + 
{1\over r \left (r^2 - 1 \right)^3} \bigg (\mu 
\text {$\beta $} (4 \left (r^2 + 3 \right) \text {a3} (r)
                r^2 + (r^2 \nonumber \\
&& \hpo - 
          1 )  (F' (r) r^6 + \text {a6}' (r) r^4 - 
         3 F' (r) r^4 - 2
              \text {a3}' (r) r^3 - 
         2 \left (r^2 - 1 \right) \text {h2} (r) r^2 \nonumber \\
&& \hpo + 2
              \left (r^2 - 1 \right) \text {j2} (r) r^2 - 
\text {a6}'(r) r^2 + 3 F' (r) r^2 - 2
              \text {a3}' (r) r - F' (r) ) 
) \bigg) \bigg ] \nonumber \\
\een
and imaginary part of the retarded Green function for 
effective action is given
  by
\ben
{1 \over  \omega} \text{Im}[G^{R^{(\text{effective})}}_{xy,xy}]&=& 
\lim_{r\ra0}\bigg [
-2 \beta -\mu \bigg({1\over \left (r^2 - 1 \right)^3} \bigg (\left (r^2 - 
          1 \right)\nonumber \\
&& \hpo (2 \Gamma  r^4 - \text {a6}' (r) r^3 - 4
              \Gamma  r^2 + 2 \text {a3}' (r) r^2 + 
         2 \left (r^2 - 1 \right) \text {h2} (r) r 
\nonumber \\
&& \hpo - 2
              \left (r^2 - 1 \right) \text {j2} (r) r 
+ \text {a6}' (r) r + 2 \Gamma + 2
              \text {a3}' (r) ) - 4 r \left (r^2 + 3 \right)
           \text {a3} (r)\bigg)   
\nonumber \\
&& \hpo - 
   r F' (r) + \frac {F' (r)} {r} \bigg)
   \beta 
\bigg ]\ .
\een
Therefore, in low frequency limit the difference between the imaginary part of
retarded Green
  function coming from this two boundary terms are given by ,
\ben
\lim_{\omega \ra 0}{1 \over  \omega} 
\text{Im}\ltb G^{R^{(\text{original})}}_{xy,xy}\rtb
- {1\over \omega}\text{Im}\ltb G^{R^{(\text{effective})}}_{xy,xy}\rtb &=&
2 \mu \ \beta \ \Gamma \ .
\een
Therefore for this general class of theory,
\be
\Gamma =0\ .
\ee

The other kind of higher derivative theory one can consider is covariant
derivatives acting on curvature tensors. In that case one can have a more
general action like (\ref{ghdacnphi}). For this kind of action the boundary
terms one gets are of the form $ \phi^{(n)}\phi^{(p)}$ (here $\phi^{(n)}$
means n-th derivative of $\phi$ with respect to $r$). Using the form of $\phi$
given in (\ref{phiform}) it can be shown that except $\phi^{(n)} \phi$ kind of
terms, other boundary terms do not contribute in low frequency limit. For
example, if we consider $C_n \phi^{(n)^2}$ term in the original action, the
the relevant boundary term which will contribute in low frequency limit is
$(-1)^{(n+1)}(C_n \phi^{(n)})^{(n-1)}\phi$. One can check that though we need
to add {\it Gibbons-Hawking} terms to make the variation of the action well
defined but most of them are zero in low frequency limit. We have checked it
for few nontrivial terms like, $\phi^{(3)^2}$ and $ \phi^{(4)^2}$ and $\Gamma$
turns out to be zero. But we expect it is true in general.

\section{Expressions for $A^{GB}$}\label{app2}

\ben
A_1^{GB}(r,k)&=&\frac{8 r^2 - \omega^2 r - 8} {4 r^3 \left (r^2 - 1 \right)} 
- {1 \over (12 (r^3
        (r^2 - 1 )^2 ))}
    ((10 {c_1} (88 r^4 - 11 \omega^2 r^3 - 176 r^2 + 13
        \omega^2 r + 88 ) \nonumber \\
&& + {c_3} (144 r^8 - 288 r^6 + 66
        \omega^2 r^5 + 232 r^4 + 25 \omega^2 r^3 - 4 (3 \omega^4
        + 44 ) r^2 + 13 \omega^2 r + 88 ) \nonumber \\
&& + {c_2} 
        (176 r^4 - 22 \omega^2 r^3 - (3 \omega^4 + 352 )
        r^2 + 26 \omega^2 r + 176 ) ) \mu)  + O(\mu^2 ) \nonumber \\
A_2^{GB}(r,k)&=&-\frac {3 (r^2 - 
      1 )} {r} \nonumber \\
&& + \frac {(10 {c_1} (13
             r^2 - 11 ) +
      2 {c_2} (2 r^4 + 17 r^2 - 9 ) + {c_3}
           (34 r^4 + 9 r^2 - 8 \omega^2 r - 3 ) )
       \mu} {r} + O(\mu^2 )  \nonumber \\
A_3^{GB}(r,k)&=&
4 ({c_2} + 4 {c_3}) r \left (r^2 - 1 \right)^2 \mu + 
 O\left (\mu^2 \right) \nonumber \\
A_4^{GB}(r,k)&=&-\frac {2 (r^2 + 
      3 )} {r^2} \nonumber \\
&& +{1\over 3 r^2(r^2 - 1)} (2 (10 {c_1} (13
             r^4 + 20 r^2 - 33 ) + {c_2} (26 r^4 + 3 \omega^2 r^3 + 40 r^2 + 3
              \omega^2 r - 66) \nonumber \\
&& + {c_3} (90 r^6 - 89 r^4 + 30 \omega^2 r^3 + 32
             r^2 + 6 \omega^2 r - 33 ) ) \mu)  + O\left (\mu^2 
\right)\nonumber \\
A_5^{GB}(r,k)&=&
-\frac {4 (r^2 - 
      1 )} {r} \nonumber \\
&& + \frac {2 (20 {c_1} (13
             r^2 - 11 ) + 
      2 {c_3} (18 r^4 + r^2 - 11 ) + {c_2} (52
             r^2 + 3 \omega^2 r - 44 )) \mu} {3 r} + 
 O\left(\mu^2 \right)\nonumber \\
A_6^{GB}(r,k)&=&
8 \left (r^2 - 1 \right) \left ({c_2} r^2 + 
   4 {c_3} r^2 + {c_2} \right)
   \mu + O\left (\mu^2 \right) \ .
\een

\section {Expressions for $A^W$}\label{app4}

\ben
A^W_1&=&
\frac {8 r^2 - \omega^2 r - 
    8} {4 r^3 \left (r^2 - 1 \right)} \nonumber \\
&& + \frac {\left (-360
           r^9 - 240 r^7 + 129 \omega^2 r^6 + 1560 r^5 - 
       300 \omega^2 r^4 + 8 \left (
\omega^4 - 120 \right) r^3 + 
       75 \omega^2 \right) \gamma } {4
        \left (r^2 - 1 \right)^2} + O\left (\mu ^2 \right) \nonumber \\
A_2^W&=&
-\frac {3 \left (r^2 - 1 \right)} {r} + 
 r \left (-419 r^6 + 668 r^4 - 24 \omega^2
         r^3 + 8 r^2 - 225 \right) \mu + 
 O\left (\mu ^2 \right)\nonumber \\
A^W_3&=&
32 r^5 \left (r^2 - 1 \right)^2 \mu + O\left (\mu^2 \right)\nonumber \\
A^W_4&=&
-\frac {2 \left (r^2 + 
      3 \right)} {r^2} - \frac {2 \left (2045 r^8 - 4185 r^6 - 26
            \omega^2 r^5 + 2140 r^4 - 2 \omega^2 r^3 + 75 r^2 - 
       75 \right) \mu
      } {r^2 - 1} + O\left (\mu^2 \right)\nonumber \\
A^W_5&=&
-\frac {4 \left (r^2 - 1 \right)} {r} - 
 4 \left (r \left (145 r^6 - 220 r^4 + 2 
\omega^2 r^3 + 75 \right) \right) \mu + 
 O\left (\mu^2 \right)\nonumber \\
A^W_6&=&
32 r^4 \left (2 r^4 - 3 r^2 + 1 \right) \mu + 
 O\left (\mu^2 \right)\ .
\een

\bc
.......................................
\ec


\begin{thebibliography}{99}


\bibitem{pss}
  G.~Policastro, D.~T.~Son and A.~O.~Starinets,
  ``The shear viscosity of strongly coupled N = 4 supersymmetric Yang-Mills
  plasma,''
  Phys.\ Rev.\ Lett.\  {\bf 87}, 081601 (2001)
  [arXiv:hep-th/0104066].

\bibitem{rev}
  D.~T.~Son and A.~O.~Starinets,
  ``Viscosity, Black Holes, and Quantum Field Theory,''
  Ann.\ Rev.\ Nucl.\ Part.\ Sci.\  {\bf 57}, 95 (2007)
  [arXiv:0704.0240 [hep-th]].

\bibitem{Kovtun:2004de}
  P.~Kovtun, D.~T.~Son and A.~O.~Starinets,
  ``Viscosity in strongly interacting quantum field theories from black hole
  physics,''
  Phys.\ Rev.\ Lett.\  {\bf 94}, 111601 (2005)
  [arXiv:hep-th/0405231].


\bibitem{Kovtun:2005ev}
  P.~K.~Kovtun and A.~O.~Starinets,
  ``Quasinormal modes and holography,''
  Phys.\ Rev.\  D {\bf 72}, 086009 (2005)
  [arXiv:hep-th/0506184].



\bibitem{ss}
  D.~T.~Son and A.~O.~Starinets,
  ``Minkowski-space correlators in AdS/CFT correspondence: Recipe and
  applications,''
  JHEP {\bf 0209}, 042 (2002)
  [arXiv:hep-th/0205051].

\bibitem{hs}
  C.~P.~Herzog and D.~T.~Son,
  ``Schwinger-Keldysh propagators from AdS/CFT correspondence,''
  JHEP {\bf 0303}, 046 (2003)
  [arXiv:hep-th/0212072].

\bibitem{janik}
  R.~A.~Janik and R.~B.~Peschanski,
  ``Asymptotic perfect fluid dynamics as a consequence of AdS/CFT,''
  Phys.\ Rev.\  D {\bf 73}, 045013 (2006)
  [arXiv:hep-th/0512162].

  R.~A.~Janik and R.~B.~Peschanski,
  ``Gauge / gravity duality and thermalization of a boost-invariant perfect
  fluid,''
  Phys.\ Rev.\  D {\bf 74}, 046007 (2006)
  [arXiv:hep-th/0606149].



\bibitem{shiraz}
  S.~Bhattacharyya, V.~E.~Hubeny, S.~Minwalla and M.~Rangamani,
  ``Nonlinear Fluid Dynamics from Gravity,''
  JHEP {\bf 0802}, 045 (2008)
  [arXiv:0712.2456 [hep-th]].



\bibitem{Bhattacharyya:2008ji} 
S.~Bhattacharyya, R.~Loganayagam,
  S.~Minwalla, S.~Nampuri, S.~P.~Trivedi and S.~R.~Wadia, 
  Fluid Dynamics from Gravity,'' arXiv:0806.0006 [hep-th].  
  = ARXIV:0806.0006;


\bibitem{Haack:2008cp}
  M.~Haack and A.~Yarom,
  ``Nonlinear viscous hydrodynamics in various dimensions using AdS/CFT,''
  arXiv:0806.4602 [hep-th].

\bibitem{Bhattacharyya:2008mz}
  S.~Bhattacharyya, R.~Loganayagam, I.~Mandal, S.~Minwalla and A.~Sharma,
  ``Conformal Nonlinear Fluid Dynamics from Gravity in Arbitrary Dimensions,''
  JHEP {\bf 0812}, 116 (2008)
  [arXiv:0809.4272 [hep-th]].






\bibitem{VanRaamsdonk:2008fp}
M.~Van~Raamsdonk, ``{Black Hole Dynamics From Atmospheric
Science},'' \href{http://www.arXiv.org/abs/arXiv:0802.3224
[hep-th]}{{\tt arXiv:0802.3224
  [hep-th]}}.


\bibitem{Loganayagam:2008is}
R.~Loganayagam, ``{Entropy Current in Conformal Hydrodynamics},''
\href{http://www.arXiv.org/abs/arXiv:0801.3701 [hep-th]}{{\tt
arXiv:0801.3701
  [hep-th]}}.


\bibitem{Heller:2008mb}
  M.~P.~Heller, P.~Surowka, R.~Loganayagam, M.~Spalinski and S.~E.~Vazquez,
  ``On a consistent AdS/CFT description of boost-invariant plasma,''
  arXiv:0805.3774 [hep-th].


\bibitem{Friess:2006kw}
J.~J. Friess, S.~S. Gubser, G.~Michalogiorgakis, and S.~S. Pufu,
``Expanding
  plasmas and quasinormal modes of anti-de Sitter black holes,'' {\em
JHEP}
  {\bf 04} (2007) 080,
\href{http://www.arXiv.org/abs/hep-th/0611005}{{\tt hep-th/0611005}}.



\bibitem{Herzog:2002fn}
C.~P. Herzog, ``The hydrodynamics of M-theory,'' {\em JHEP} {\bf 12}
(2002)
  026,
\href{http://www.arXiv.org/abs/hep-th/0210126}{{\tt hep-th/0210126}}.

\bibitem{Policastro:2002tn}
G.~Policastro, D.~T. Son, and A.~O. Starinets, ``From AdS/CFT
correspondence to
  hydrodynamics. II: Sound waves,'' {\em JHEP} {\bf 12} (2002) 054,
\href{http://www.arXiv.org/abs/hep-th/0210220}{{\tt hep-th/0210220}}.



\bibitem{Gupta:2008th}
  R.~K.~Gupta and A.~Mukhopadhyay,
  ``On the universal hydrodynamics of strongly coupled CFTs with gravity
  duals,''
  arXiv:0810.4851 [hep-th].






\bibitem{Herzog:2003ke}
C.~P. Herzog, ``The sound of M-theory,'' {\em Phys. Rev.} {\bf D68}
(2003)
  024013,
\href{http://www.arXiv.org/abs/hep-th/0302086}{{\tt hep-th/0302086}}.



\bibitem{Buchel:2008ae}
  A.~Buchel, R.~C.~Myers, M.~F.~Paulos and A.~Sinha,
  ``Universal holographic hydrodynamics at finite coupling,''
  arXiv:0808.1837 [hep-th].


\bibitem{Buchel:2008vz}
  A.~Buchel, R.~C.~Myers and A.~Sinha,
  ``Beyond eta/s = 1/4pi,''
  arXiv:0812.2521 [hep-th].



\bibitem{Banerjee:2008th}
  N.~Banerjee, J.~Bhattacharya, S.~Bhattacharyya, S.~Dutta, 
  R.~Loganayagam and P.~Surowka,
  ``Hydrodynamics from charged black branes,''
  arXiv:0809.2596 [hep-th].





\bibitem{Erdmenger:2008rm}
  J.~Erdmenger, M.~Haack, M.~Kaminski and A.~Yarom,
  ``Fluid dynamics of R-charged black holes,''
  arXiv:0809.2488 [hep-th].





\bibitem{Myers:2008yi}
  R.~C.~Myers, M.~F.~Paulos and A.~Sinha,
  ``Quantum corrections to eta/s,''
  arXiv:0806.2156 [hep-th].













\bibitem{Buchel:2008kd}
  A.~Buchel and M.~Paulos,
  ``Second order hydrodynamics of a CFT plasma from boost invariant
  expansion,''
  arXiv:0808.1601 [hep-th].


\bibitem{Shuryak:2003xe}
E.~Shuryak, ``Why does the quark gluon plasma at RHIC behave as a nearly
ideal
  fluid?,'' {\em Prog. Part. Nucl. Phys.} {\bf 53} (2004) 273--303,
\href{http://www.arXiv.org/abs/hep-ph/0312227}{{\tt hep-ph/0312227}}.

\bibitem{Shuryak:2004cy}
E.~V. Shuryak, ``What RHIC experiments and theory tell us about
properties of
  quark-gluon plasma?,'' {\em Nucl. Phys.} {\bf A750} (2005) 64--83,
\href{http://www.arXiv.org/abs/hep-ph/0405066}{{\tt hep-ph/0405066}}.

\bibitem{Shuryak:2006se}
E.~V. Shuryak, ``Strongly coupled quark-gluon plasma: The status
report,''
\href{http://www.arXiv.org/abs/hep-ph/0608177}{{\tt hep-ph/0608177}}.







\bibitem{brss}
  R.~Baier, P.~Romatschke, D.~T.~Son, A.~O.~Starinets and M.~A.~Stephanov,
  ``Relativistic viscous hydrodynamics, conformal invariance, and holography,''
  JHEP {\bf 0804}, 100 (2008)
  [arXiv:0712.2451 [hep-th]].






\bibitem{no}
  M.~Natsuume and T.~Okamura,
  ``Causal hydrodynamics of gauge theory plasmas from AdS/CFT duality,''
  Phys.\ Rev.\  D {\bf 77}, 066014 (2008)
  [Erratum-ibid.\  D {\bf 78}, 089902 (2008)]
  [arXiv:0712.2916 [hep-th]].

\bibitem{gy}
  S.~S.~Gubser and A.~Yarom,
  ``Linearized hydrodynamics from probe-sources in the gauge-string duality,''
  arXiv:0803.0081 [hep-th].



\bibitem{buchel}
  A.~Buchel, J.~T.~Liu and A.~O.~Starinets,
  ``Coupling constant dependence of the shear viscosity in N=4 supersymmetric
  Yang-Mills theory,''
  Nucl.\ Phys.\  B {\bf 707}, 56 (2005)
  [arXiv:hep-th/0406264].



%


\bibitem{kats}
  Y.~Kats and P.~Petrov,
  ``Effect of curvature squared corrections in AdS on the viscosity of the dual
  gauge theory,''
  arXiv:0712.0743 [hep-th].





\bibitem{myers1}
  M.~Brigante, H.~Liu, R.~C.~Myers, S.~Shenker and S.~Yaida,
  ``Viscosity Bound Violation in Higher Derivative Gravity,''
  Phys.\ Rev.\  D {\bf 77}, 126006 (2008)
  [arXiv:0712.0805 [hep-th]].








\bibitem{ramy1}
  R.~Brustein and A.~J.~M.~Medved,
  ``The ratio of shear viscosity to entropy density in generalized theories of
  gravity,''
  arXiv:0808.3498 [hep-th].


\bibitem{ramy2}
  R.~Brustein and A.~J.~M.~Medved,
  ``The shear diffusion coefficient for generalized theories of gravity,''
  arXiv:0810.2193 [hep-th].




\bibitem{liu}
  N.~Iqbal and H.~Liu,
  ``Universality of the hydrodynamic limit in AdS/CFT and the membrane
  paradigm,''
  arXiv:0809.3808 [hep-th].

\bibitem{stretch}
  P.~Kovtun, D.~T.~Son and A.~O.~Starinets,
  ``Holography and hydrodynamics: Diffusion on stretched horizons,''
  JHEP {\bf 0310}, 064 (2003)
  [arXiv:hep-th/0309213].



\bibitem{cai1}
  R.~G.~Cai, Z.~Y.~Nie and Y.~W.~Sun,
  ``Shear Viscosity from Effective Couplings of Gravitons,''
  Phys.\ Rev.\  D {\bf 78}, 126007 (2008)
  [arXiv:0811.1665 [hep-th]].

  R.~G.~Cai, Z.~Y.~Nie, N.~Ohta and Y.~W.~Sun,
  ``Shear Viscosity from Gauss-Bonnet Gravity with a Dilaton Coupling,''
  arXiv:0901.1421 [hep-th].






\bibitem{buchel2}

  A.~Buchel,
 ``Shear viscosity of boost invariant plasma at finite coupling,''
  Nucl.\ Phys.\  B {\bf 802}, 281 (2008)
  [arXiv:0801.4421 [hep-th]].



  A.~Buchel,
  ``Resolving disagreement for eta/s in a CFT plasma at finite coupling,''
  Nucl.\ Phys.\  B {\bf 803}, 166 (2008)
  [arXiv:0805.2683 [hep-th]].






\bibitem{dutta}
  S.~Dutta,
  ``Higher Derivative Corrections to Locally Black Brane Metrics,''
  JHEP {\bf 0805}, 082 (2008)
  [arXiv:0804.2453 [hep-th]].


\bibitem{gks}
  S.~S.~Gubser, I.~R.~Klebanov and A.~A.~Tseytlin,
  ``Coupling constant dependence in the thermodynamics of N = 4  supersymmetric
  Yang-Mills theory,''
  Nucl.\ Phys.\  B {\bf 534}, 202 (1998)
  [arXiv:hep-th/9805156].






\bibitem{dg}
  S.~Dutta and R.~Gopakumar,
  ``On Euclidean and noetherian entropies in AdS space,''
  Phys.\ Rev.\  D {\bf 74}, 044007 (2006)
  [arXiv:hep-th/0604070].

\bibitem{wip}
Work in progress with R. Brustein and D. Gorbons.


























































\end{thebibliography}
\end{document}